# Ultralow-loss spiral resonators for precise LiDAR

OSAMA TERRA[1,2], WARREN JIN[1,3], HUSSEIN KOTB[4], JOEL GUO[1], AND JOHN E. BOWERS[1,*]

[1]*Department of Electrical and Computer Engineering, University of California, Santa Barbara, CA, USA*
[2]*Primary Length and Laser Technology lab., National Institute of Standards, Giza, Egypt*
[3]*Anello Photonics, Santa Clara, CA, USA*
[4]*Department of Electronics and Electrical Communication Engineering, Faculty of Engineering, Ain Shams University, Cairo, Egypt.*
*\*bowers@ece.ucsb.edu*

**Abstract:** Swept laser interferometry is an extremely powerful solution embedded in several recent technologies such as absolute distance measurement, light detection and ranging, optical frequency domain reflectometry, optical coherence tomography, microresonator characterization, and gas spectroscopy. Nonlinearity in the optical frequency sweeping of tunable lasers is a fatal drawback in gaining the expected outcome from these technologies. Here, we introduce an on-chip, millimeter-scale, 7 m spiral resonator that is made of ultra-low loss $Si_3N_4$ to act as a frequency ruler for correction of the tunable lasers sweeping nonlinearities. The sharp 2 MHz frequency lines of the 85 M Q high-quality resonator and the narrow-spaced 25.57 MHz frequency ticks of the 7 m spiral allow unprecedented precise nonlinearity correction on an integrated photonics platform. Accurate measurements of the ruler's frequency spacing, linewidth, and temperature and wavelength sensitivities of the frequency ticks are performed here to demonstrate the quality of the frequency ruler. In addition, the spiral resonator is implemented in an FMCW LiDAR experiment to demonstrate a potential application of the proposed on-chip frequency ruler.

## 1. Introduction

Swept laser interferometry (SLI) is a common technique in many important applications such as distance metrology [1, 2], distributed fiber sensing using Optical Frequency Domain Reflectometry (OFDR) [3], autonomous vehicle guidance systems using Frequency-Modulated Continuous-Wave Light Detection and Ranging (FMCW-LiDAR) [4, 5], Gas Spectroscopy to detect gas pressure using the broadening in linewidth [6, 7], integrated photonics for characterizing microresonators [8], and in eye disease identification using Swept-Source Optical Coherence Tomography (SS-OCT) [9]. However, wavelength tuning nonlinearity is considered one of the most challenging obstacles towards achieving the required measurement precision using SLI technology [10, 11].

Ideally, an optical frequency ruler with close spacing, exact, and stable frequency spacing can provide a reference for the correction of SLI wavelength scanning nonlinearities. The closer the spacing between the frequency lines, the higher the sampling rate of the reference frequency ruler and hence the better its nonlinearity correction resolution. Several schemes have been proposed to be used as a reference. These schemes include reference Fabry Perot interferometers etalons (FP), Mach-Zehnder interferometers, and fiber ring resonators [12-14]. Each of these techniques has its pros and cons. Although FP cavities provide accurate and sharp transmission lines, the cavity length needs to be extremely large to achieve higher correction sampling points, which is not suitable for integrated photonics systems and introduces vibration and temperature sensitivity. Mach-Zehnder Interferometers (MZI) provide a simple solution for correcting the nonlinearities in frequency sweeping systems and can be easily integrated into silicon photonics platforms [4, 15]. However, detecting the peaks of the sine waves generated from the MZI during the frequency sweeping is not precise, which imposes the use of complex and time-consuming phase detection algorithms to improve the correction precision [16]. Recently, a fiber ring scheme has been proposed to generate equidistant narrow sharp transmission lines to act as a reference frequency marker for correcting sweeping nonlinearities [14], but there is not yet a suitable solution for the fast-growing integrated photonics platforms and reduced environmental sensitivity is possible with solid-state waveguides.

With the recent advances in integrated photonic components including the development of heterogeneously integrated narrow linewidth lasers and the mode-hop free tunable lasers [17, 18], there is a crucial need for an on-chip frequency referencing to correct laser sweeping nonlinearities. Another major advancement in integrated photonics platforms is the development of ultra-low loss silicon nitride waveguides and resonators that can reach loss below 0.1 dB/m [19, 20], which will lead in turn to the development of high-quality passive photonics structures on chip.

In this paper, a Spiral Resonator (SPR) with a spiral length of 7 m that is fabricated from ultra-low loss silicon nitride is introduced to act as a narrow-spacing frequency ruler for correcting the frequency sweeping nonlinearities in tunable lasers. The SPR is designed in a small footprint and reproducible design which would greatly expand its outreach and applications of the SLI technique in the integrated photonics platforms. Full characterization of the SPR is made for the parameters that contribute to the accuracy of the frequency ruler such as the frequency spacing between the frequency lines (free-spectral range), resonator quality factor, resonance linewidth, and the dependence of the line-spacing on the temperature variations and wavelength change. Finally, a practical example of using the SPR for the FMCW-LiDAR application will be given to demonstrate the potential of the SPR to enhance the measurement accuracy of these systems.

## 2. Device description

Ultra-low loss waveguides are required for achieving high-quality factor resonators, especially for meter-scale resonators. Scattering from the rough waveguide walls is the main reason for loss; therefore, waveguides with a large aspect ratio (width/thickness>>1) are desirable for achieving lower propagation loss [21]. By increasing the width of the waveguide, the loss due to the sidewall scattering is reduced and the light propagation inside the ring resonator will be in the whispering gallery mode; and by decreasing the thickness of the waveguide, the scattering loss is minimized due to the lower modal confinement [21]. Fig. 1a shows the cross-section of the SPR chip, which shows the $Si_3N_4$ waveguide width of 10 µm and thickness of 100 nm. It provides a bending radius greater than 700 µm to avoid the radiation loss caused by a small bending radius of curvature [21]. The upper and lower $SiO_2$ claddings have thicknesses of 2 and 14.5 µm. The cavity of the SPR, having its cross-section area in Fig. 1a, is designed by connecting two multimode Archimedean spirals with an S-shape bend at the center [22], and connecting the other ends of the two spirals at the outer part. The spiral waveguide is adiabatically tapered from 10 µm to 2.8 µm to match the waveguide width of the S-bent at the center to allow the propagation of only the fundamental mode while damping the higher order modes. Fig. 1b shows a simple schematic of the SPR, which shows the structure of the SPR, input port, and drop port. Two similar directional couplers supporting single TE-mode waveguides are used to construct the input and output ports of the ring resonator. As shown in Fig. 1b, the light is coupled to the spiral resonator at the input port and is coupled out either from the through port or the drop port. The gap width of the directional couplers is tapered to minimize the coupling losses [23]. Fig. 1c shows a photograph of the SPR compared to one US cent to show the small footprint of the resonator. The fundamental TE-mode profile of the 10 µm×100 nm $Si_3N_4$ waveguide is plotted in Fig. 1d to show the low confinement of the mode due to this thickness. The intrinsic and the loaded quality factors measured for the SPR are measured to be around 75 to 85 Million at 1550 nm as shown in Fig. 1e.

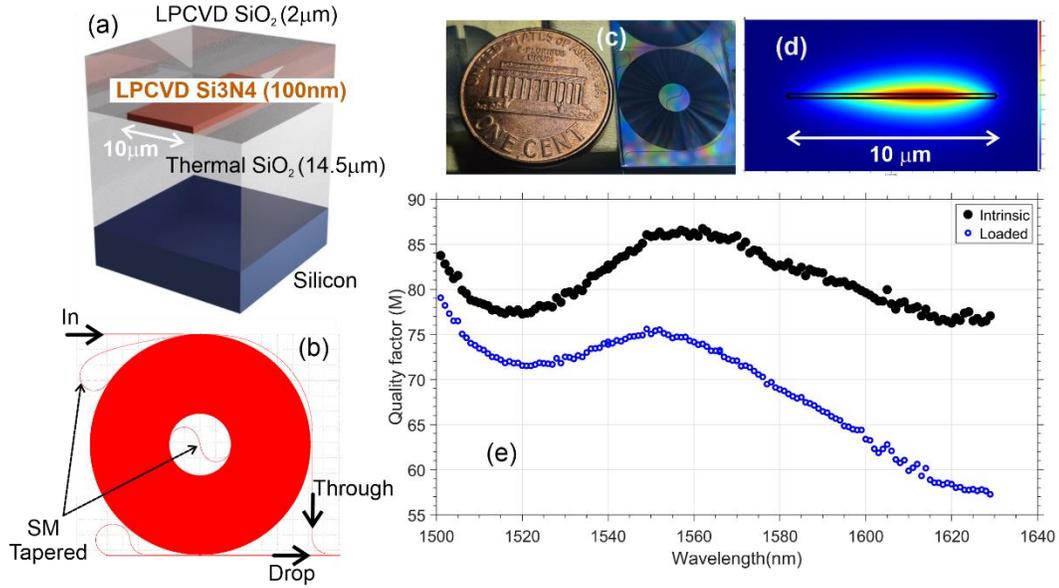

Fig. 1. Ultralow loss $Si_3N_4$ 7m spiral resonator, (a) cross-sectional view that shows that the thickness of the waveguide is 100 nm, (b) the spiral resonator layout (through and drop port are shown), (d) the mode-profile from the propagating mode inside the 10 μm wide and 100 nm thick resonator, (e) photograph of the spiral resonator in comparison with US 1 cent, (f) intrinsic and loaded quality factors of the spiral resonator.

## 3. Laser sweeping nonlinearity correction

SPR is used to correct the laser sweeping nonlinearity while measuring the Free-Spectral Range (FSR) of a fiber-based unbalanced MZI. Measuring the FSR beat frequency is a common method in FMCW-LiDAR, OFDR, and other SLI experiments. A simple schematic for the experiment is depicted in Fig 2a, the output from a tunable laser is split between two ports to have 90% of the light to the SPR and the other 10% to the unbalanced MZI. Edge coupling to the SPR chip is made by cleaving the fiber end and using index-matching gel to facilitate coupling and avoid reflections from the air gap. A normal single-mode fiber with a polarization rotator is used here for the experiment to be compatible with other components in the laboratory. The transmission peaks from the SPR and the interference fringes from the MZI are detected using two photodetectors and acquired by two channels of an oscilloscope. The laser is set to have a sweeping period of 1 nm and a sweeping speed of 5 nm/s, which means that the oscilloscope will acquire around 4850 SPR peaks at 200 ms. In Fig. 2b, two zoomed oscilloscope traces for the MZI (upper) and the SPR (lower) are demonstrated before applying the correction. Two similar double arrow lines are drawn at the MZI trace to demonstrate the slight change in the periods; however, the change in the period can be more severe in other cases. Even with this slight change the Fast Fourier Transform (FFT) of the varying sinusoidal trace shows great broadening in the FSR peak as shown in Fig. 2c. After applying a program to correct the laser sweeping nonlinearity the broadened FSR peak is reduced to a narrow peak the exact FSR of the unbalanced MZI as demonstrated in Fig. 2d. The program is based on referencing the MZI trace to detected sharp peaks with very well-known frequency spacings to resample the time scale of the oscilloscope into frequency scale. The spacing between the SPR peaks is precisely measured by the swept frequency modulation technique demonstrated in the next section to be exactly 25.566 MHz. The corrected FSR of the MZI is 40.31 MHz with a shift of 2.25 MHz from the uncorrected FSR. This correction is crucial to several applications such as in FMCW-LiDAR as will be demonstrated later in the paper.

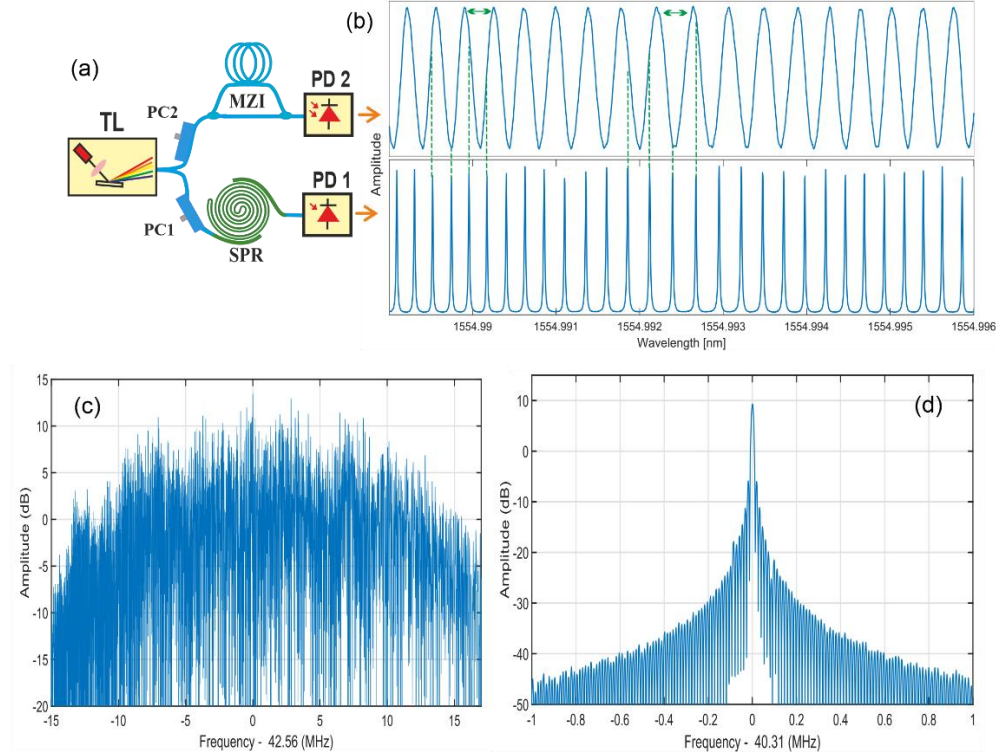

Fig. 2. Correction of laser sweeping nonlinearity using the drop port of the spiral resonator (SPR) (a) the setup implemented for correction of the sweeping nonlinearity of a tunable laser (TL) during measurement of the FSR of a Mach-Zehnder interferometer (MZI) using the SPR, PD: photodetector, PC: polarization controller, (b) Oscilloscope traces for the MZI (upper) at CH1 and the spiral resonator (lower) at CH2 that acquired during laser is sweeping (parts of the traces are shown), (c) FFT of the swept trace before nonlinearity correction and (d) FFT of the swept trace after applying the nonlinearity correction. The center frequency represents the FSR of the interferometer.

## 4.  FSR and Linewidth measurement of the SPR

Several techniques have been developed for the measurement of the FSR of high-quality factor microresonators and Fabry-Perot interferometers. Among those techniques are the frequency difference technique [24], the null method technique [25], and the phase-modulated sideband technique [26]. The last method is commonly used for the measurement of the FSR and relies on referencing the measurement of the oscilloscope time trace to the frequency spacing between the carrier and a modulated sideband. Although this method is commonly used to measure the FSR of micro-resonators, its accuracy is still not sufficient since it still depends on the oscilloscope resolution. Here, a method is introduced to measure the resonator's FSR and linewidth by locking the laser frequency to one of the resonance peaks while sweeping its modulation sidebands on the adjacent resonances, which is schematically demonstrated in Fig. 3a. The heterodyne interference between the laser and its constantly swept side-bands gives an amplitude-dependent spectrum on the electrical spectrum analyzer centered at the FSR frequency as depicted in Fig. 3b. By fitting a Lorentzian curve to the spectrum, the linewidth and the FSR are obtained. The measurement setup consists mainly of a narrow linewidth laser, a frequency locking system to lock the laser frequency to the central resonance, and a frequency modulation system as demonstrated in Fig. 3d. and explained in more detail in the methods section of this paper.

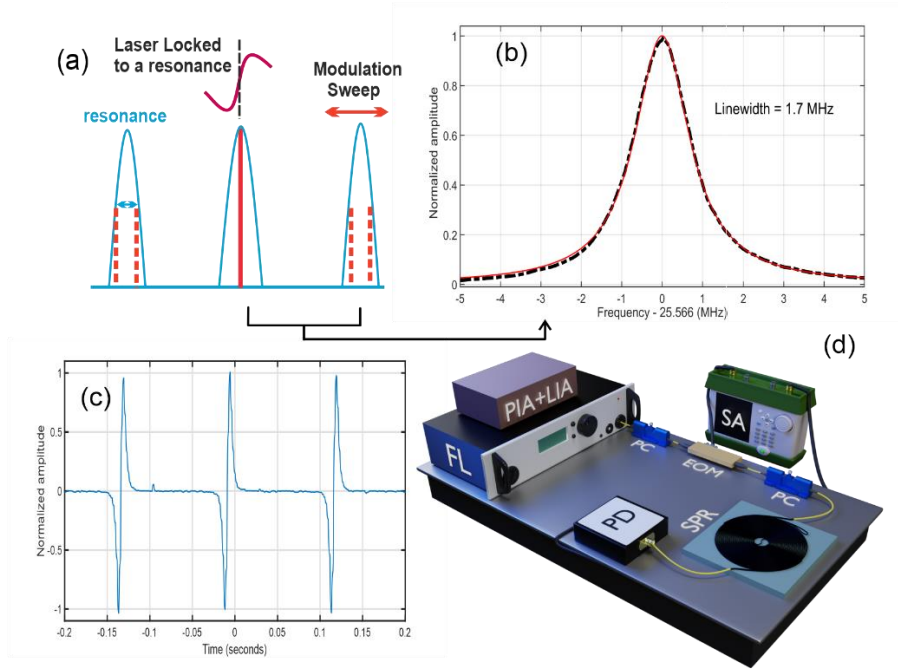

Fig. 3. (a) Concept of the modulated swept method to determine the FSR and resonance linewidth of the spiral resonator, the laser is locked to the central peak, while the modulation sideband is swept over the adjacent resonances, the heterodyne interference between the central peak and the sideband is used to deduce the FSR (b) Transmission spectrum from the spiral resonator. red line: Lorentzian curve fitting (c) error signal for three sweeps over the resonance, which shows the signal used for locking the laser to the resonance (d) FSR measurement setup based on the modulated-wave sweeping method, FL: fiber laser, EOM: Electro-Optic Modulator, PC: polarization controller, SPR: Spiral resonator, PD: Photodetector, LIA: Lock-in-Amplifier, PID: Servo-controller, SA: Spectrum analyzer with frequency tracking generator.

The laser is locked to the central resonance by slightly dithering the frequency of the laser to generate a dispersion-like signal, which is shown in Fig. 3c. Proportional-Integral-Derivative (PID) uses the generated dispersion-like signal to lock the laser frequency to the resonance peak. The laser is additionally amplitude modulated using an Electro-Optic Modulator (EOM) such that the modulation sidebands match the peaks of the next resonances, the heterodyne beat between the laser central frequency and its sidebands shows maximum amplitude for maximum transmission of the sidebands. On the other hand, when the modulation frequency is detuned from the FSR center frequency, the amplitude of the interference signal decreases due to the decrease in the sideband transmission power. By sweeping the EOM modulation frequency around the side-band frequency using the tracking generator of a spectrum analyzer, while acquiring the beat on the same spectrum analyzer, a transmission peak will be displayed which holds the FSR frequency and resonance linewidth, see Fig. 3b.

## 5. Chromatic dispersion and thermo-optic coefficients

The FSR of the SPR is measured at a temperature of 23 °C and at a wavelength of 1556.2 nm. However, it is expected that the conditions during the applications will deviate from the stated conditions; therefore, it is important to consider the change in FSR when deviating from these conditions. Two parameters are measured here to allow the correction of the FSR; namely, the chromatic dispersion and the thermo-optic coefficients.

-*Chromatic Dispersion:* Several methods have been proposed for measuring microresonator dispersion, which are summarized in a review article by S. Fujii et al. [27]. Here, a slightly modified method based on the calibrated reference Mach-Zehnder interferometer method is presented in the paper. A Fiber-Ring Resonator (FRR) is designed, fabricated, and calibrated to be used as a reference for dispersion measurement. The FRR is designed to have resonances that are similar in FSR and linewidth to that of the SPR, but with much lower dispersion due to the much lower dispersion of the silica-based fibers when compared to that of the $Si_3N_4$ waveguides. As demonstrated in Fig. 4a, the light from the tunable laser is divided between the SPR and the reference FRR. The wavelength of the tunable laser is swept over steps of 10 nm, while recording the acquired peaks on an oscilloscope. After correction of sweeping nonlinearities using the calibrated FSR of the FRR, the SPR integrated dispersion parameter is calculated for the 10 nm wavelength range from the change in the FSR from that of the center wavelength, and plotted against the resonance number as shown in Fig. 4b. The second order dispersion is calculated from the parabolic fitting of the integrated dispersion. Fig. 4c depicts the group velocity dispersion (GVD) and the chromatic dispersion coefficient (D) over the wavelengths from 1500 nm to 1620 nm. GVD and D are calculated from the change of the FSR with optical frequency as described in detail in the methods section.

-*Thermo-optic Coefficient:* As the SPR temperature increases, its refractive index will increase which will lead to an increase in the optical length of the SPR and a decrease in its absolute resonance frequency. Provided that the laser frequency is stable during the course of the measurement and the FSR is well known from the measurement in Section 3, the decrease in the resonance frequency can be calculated easily by counting the number of resonance peaks on an oscilloscope as the SPR temperature changes (providing that FSR change is negligible to the absolute resonance frequency change). Therefore, the thermo-optic coefficient ($\frac{dn}{dT} = n \frac{dv}{dT}$) is calculated to be $1.08 \times 10^{-5}$ /K from the frequency change ($dv$) that corresponds to a temperature change ($dT$) as demonstrated in Fig 4d. This thermo-optic coefficient is very near to that of the silicon oxide due to the low confinement of the mode [28].

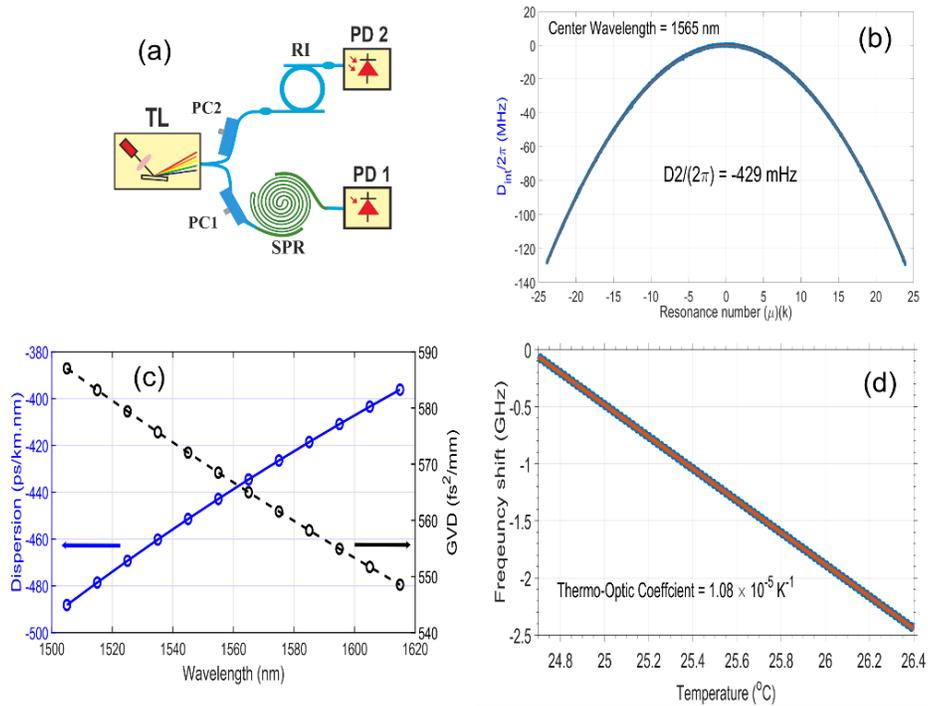

Fig. 4. (a) Chromatic dispersion measurement setup with reference to the calibrated FRR, TL: tunable laser, PC: polarization controller, PD: photodetector, FRR: calibrated fiber ring resonator, SPR: spiral resonator under measurement, scope: large record length oscilloscope (6.25 M) (b) the integrated dispersion of the spiral resonator over a wavelength range of 10 nm centered at 1565 nm and the calculated dispersion parameters (c) dispersion coefficient (left) and group velocity dispersion (right) calculated from the integrated dispersions measured at each center wavelength (d) a resonance frequency shift introduced as per SPR temperature change of 1.6 K that is used to calculate the thermo-optic coefficient.

## 6. Application example: FMCW-LiDAR

Precise ranging is important for several applications including metrology, military, large-scale manufacturing, and autonomous vehicle driving. Several techniques have been commonly used for ranging applications such as the time-of-flight, correlation of femtosecond pulses, mode-locking, and frequency sweeping interferometry [29-33]. Among these techniques, frequency sweeping interferometry allows precise ranging with a cost-effective setup; however, measuring the frequency sweeping range is considered a challenge. With the recent advancement of integrated photonic circuits and integrated tunable lasers, the need for on-chip frequency reference has become of crucial importance. Contrary to the currently used on-chip MZI with a sine-like signal, the sharp slopes and peaks of the spiral resonances determine the correction locations more precisely. Here, a round-trip distance of up to 40 m is measured using the frequency sweeping interferometry. The setup consists of a tunable laser with two output ports, the first is directed to the SPR for nonlinearity correction and another port is connected to a collimator. As depicted in Fig. 5a, the collimator sends the collimated beam from a tunable laser to a Retro-Reflector which is placed at the target location to which the distance needs to be measured.

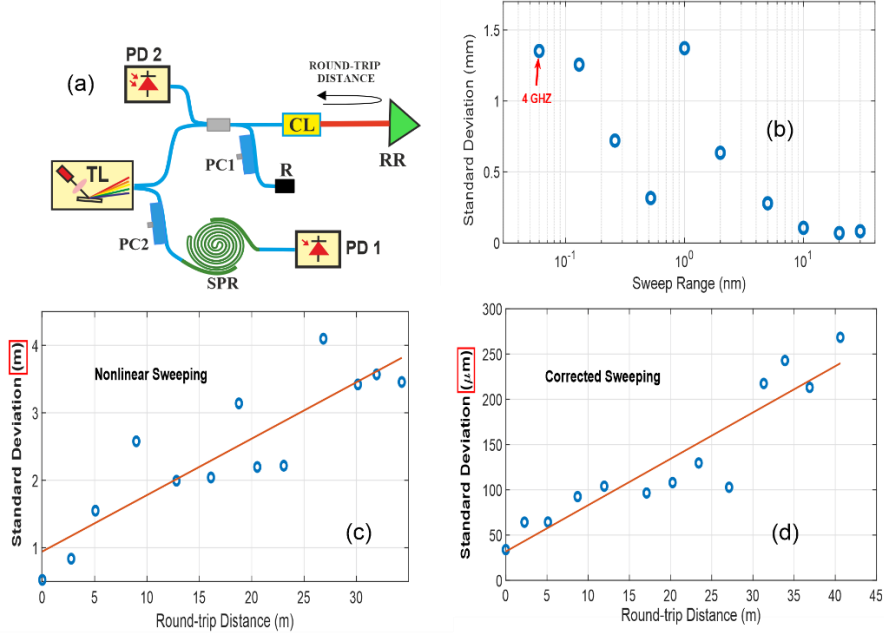

Fig. 5 (a) LiDAR experiment of round-trip distance of up to 40 m. PC: Polarization Controller, PD: Photodetector, CR: Circulator, CL: Collimator, RR: Retro-Reflector, SPR: 7 m spiral resonator, (b) standard deviation of measurement of 8 m with different laser sweeping ranges from (1 nm to 30 nm) (c) standard deviation of the measured round-trip distances up to 40 m by sweeping the wavelength of a tunable laser over 10 nm without correcting the sweeping nonlinearity (d) after correcting the sweeping nonlinearity with the SPR which shows enhancement of 4 orders of magnitude in precision.

The reflected beam from the reflector is made to interfere with the light reflected from a reference arm at the photodetector (PD2). As the laser wavelength of the tunable laser is swept, a sinusoidal signal is generated at the second channel of the oscilloscope and the reference peaks of the SPR are generated at the first channel with similar curves to those shown in Fig. 2b. After using the calibrated SPR to correct for the sweeping nonlinearity of the tunable laser, the round-trip distance is calculated by taking the FFT of the corrected distance interferometer signal. For each distance, at least 10 measurements are taken so that the standard deviation can be calculated. Fig. 5c shows the standard deviation of the round-trip distances up to around 40 m without using the SPR trace for nonlinearity correction for 10 nm sweeping range, which reaches around 4 m for the longest measured distance. In contrast, when applying the nonlinearity correction using the SPR trace, the standard deviation reaches less than 270 μm for the longest measured distance of 40 m as shown in Fig. 5d, which means an enhancement of 4 orders of magnitude in precision. Fig. 5b demonstrates the degradation of the precision with the decrease of the frequency sweeping range, which reflects that even with a range of 4 GHz, the precision is still under 1.5 mm, unlike other publications that use the MZI or ring resonators as a reference where the precision in the few cm levels [4, 36-38]. Table 1 demonstrates a comparison between the precision achieved in this work and the precision achieved by other works that use integrated MZI or ring resonators as a reference for mitigating the laser sweeping nonlinearity.

Table 1: Comparison between the precision achieved by FMCW- LiDARs referenced to the SPR in this work referenced to MZI in other state-of-art publications

|  | SWEEPING RANGE | PRECISION | REFERENCE | RANGE REPORTED |
|---|---|---|---|---|
| THIS WORK | 4 GHz- 10 nm | 100 μm -1 mm | Spiral Resonator | 40 m (round-trip) (extendable) |
| [36] | 1.2 GHz | 12.5 cm | Ring resonator | 10 m |
| [37] | 4.2 GHz | 4.6 cm | Mach Zehnder | 31 m |
| [38] | 1.5 GHz | 16.7 cm | Mach Zehnder | 75 m |
| [4] | 525 MHz | 28 cm | Mach Zehnder | 60 m |

In a work published by G. Lihachev [36], they locked a laser to a ring resonator with an attached actuator for active control of the resonator length. They tune the length of the ring resonator to linearly sweep the laser over only 1.2 GHz limited by the locking range, which enabled them to obtain 12.5 cm precision in measuring 10 m distance. In [37], a short delay-line self-heterodyne Mach-Zehnder interferometer is used to generate a chirped frequency while sweeping the laser over up to a frequency range of 4.2 GHz. Hilbert transformation of the chirped frequency is performed to deduce the instantaneous frequency. They take the previous distortions to calculate a function to correct for the next sweeps. An iterative learning algorithm is performed of up to 15 iterations before the real measurement to obtain a linear FMCW. A LiDAR experiment is performed using this technique for ranging up to 31 m with 4.6 cm precision. The same technique has been used by K. Sayyah et al. [38] to correct for the laser sweeping nonlinearity to enable 16.7 cm measurement accuracy for the 75 m range. In another paper by A. Martin et al [39], they used a Mach Zender interferometer with an external 5 m fiber delay line to have a detectable FFT signal from a laser sweeping range of 525 MHz; however, they reported a precision of 28 cm in measuring a distance of 60 m. Here, we used the spiral resonator to measure a round-trip distance of 40 m with a precision of 100 μm when the laser sweeping range was 10 nm and a precision of 1.5 mm when the sweeping range was 4 GHz which demonstrates the significant improvement in precision when the 7 m spiral resonator is used. This range depends only on the space available inside the laboratory and can be extended. The extension of the range depends mainly on the linewidth of the laser used since narrower linewidths assure high coherence ranges. For the work published by C. Xiang et al [17] for on-chip lasers, Lorentzian linewidths down a few Hz could be achieved.

## 7. Methods

### 7.1 Device Fabrication

The 7-meter spiral resonator was fabricated on a 200-mm diameter silicon substrate at a commercial CMOS foundry [21]. The 100-nm-thick stoichiometric $Si_3N_4$ waveguide layer was deposited via LPCVD on top of a 14.5 µm thick thermal oxide. The top cladding consists of a 2 µm thick layer of $SiO_2$ deposited by LPCVD using a TEOS. The wafers are annealed at 1,150 degrees C for over 20 hours to drive out residual hydrogen in the LPCVD-deposited layers, drastically reducing the absorption loss.

*7.2 Quality factor measurement*

Measuring the quality factor of the SPR is performed by sweeping the tunable laser (Santec TSL510) over steps of 1 nm each while detecting the transmission from the SPR. To correct for the laser sweeping nonlinearity, a ring fiber resonator is designed to have similar FSR and resonance linewidth to that of the SPR. An Oscilloscope (Keysight-DSO3012A) with 250 k of record length is used to acquire the transmission peaks (around 5-6 k peaks in 1 nm) of each of the SPR and the ring resonator. A computer program uses the fiber ring peaks with the known spacing on the first oscilloscope channel to convert its time scale into a uniform frequency scale. A fitting is applied to each SPR resonance peak to calculate the parameters needed to calculate the intrinsic and the loaded quality factors.

*7.3 Free-Spectral Range measurement*

An NKT photonics (Koheras-Adjustik) single-frequency fiber laser is coupled to the SPR using a normal single-mode fiber with index-matching gel to avoid reflections. The light from the drop port is coupled to a single-mode fiber connected to the 200 MHz bandwidth photodetector. The laser frequency is locked to one of the SPR resonances by modulating the fiber laser piezo input using a lock-in-amplifier (SRS MODEL SR850) to generate the dispersion-like error signal, shown in Fig 2c. The error signal is needed to lock the laser to the resonance using a laser servo-controller (Newport-LB1005) which is connected to a function generator (SRS-DS345) to generate the sawtooth signal for sweeping the laser over the resonance. A spectrum analyzer (Anritu MS 2712E) with a sweeping generator is used to drive an Electro-opics modulator so that the generated sidebands are swept over the next resonances. The beat between the center frequency and the sideband is detected on the photodetector as shown in Fig. 3b. The center of this spectrum represents the FSR of the SPR and the linewidth represents the resonance linewidth.

7.4 Chromatic dispersion measurement

The fiber ring resonator is used as a reference for measuring the SPR chromatic dispersion since the chromatic dispersion of the silica-based fiber is well-known from previous measurements using the optoelectronic-oscillation technique by the same authors [35]. In addition, it is expected that the silica-based fiber has much lower chromatic dispersion than that of the silicon nitride waveguides. The FSR of the fiber ring resonator is measured using the modulation sweeping technique discussed previously at a wavelength of 1556 nm to 27.3 MHz. The FSR is then corrected for the chromatic dispersion at each other wavelength from 1520 nm to 1630 nm using the values obtained from the fiber chromatic dispersion measurement. The laser is swept over 10 nm, to ensure a wide enough range to observe the change in chromatic dispersion change, while recording the peaks on two channels of an oscilloscope. The oscilloscope (Yokogawa DLM 2022) should have a large record length (6.25 M) to enable the acquisition of a large number of peaks at the 10 nm sweeping range (around 50 k peaks) with sufficiently high resolution. A computer program is used to compare the spacing between the resonances of both the SPR and the calibrated SRR to deduce the deviation of the SPR resonance spacing from that of the SRR. The integrated dispersion parameter is calculated for the 10 nm wavelength range from the change in the FSR from that of the center wavelength using the relation $D_{int} = (\omega_\mu - \omega_o) - \mu D_1$, where $\omega_o$ is the center frequency, $\mu$ is the number of resonances from the center frequency, $D_1 = 2\pi FSR_o$ and $FSR_o$ is the free-spectral range of the SPR at the center. The integrated dispersion ($D_{int}(\mu)$) is plotted against the resonance number ($\mu$) as shown in Fig. 4b. The second order dispersion is calculated from the parabolic fitting of $D_{int}(\mu)$ using the relation of $D_{int}(\mu)$ using the relation: $D_{int}(\mu) = 0.5 D_2 \mu^2$. The group-velocity dispersion (GVD) ($\beta_2$) is calculated from the experimental results from the variation of the free spectral range ($FSR(\omega)$) with the frequency $\beta_2 = \frac{1}{L}\frac{d}{d\omega}\left(\frac{1}{FSR(\omega)}\right)$, and $L$ is the length of the spiral. Then, the SPR GVD ($\beta_{2\_SPR}$) is extracted by subtracting the fiber GVD from the calculated values. The chromatic dispersion coefficient ($D$) can be calculated from $D = -\frac{2\pi c \beta_{2_{SPR}}(\omega_n)}{(\lambda_n)^2}$, where $c$ is the speed of light in free space. Hence, $D$ is calculated at the center angular frequency ($\omega_n$), corresponding to central wavelength ($\lambda_n$), of each 10 nm wavelength span over wavelength range from 1500 nm to 1620 nm.

**Funding:** This research was funded by DARPA MTO under a Gryphon grant and by Anello Photonics.

**Acknowledgments.** We acknowledge Mario Paniccia and Avi Feshali for assistance in managing the foundry run, O. Terra acknowledges the support of the Fulbright scholar program.

**Disclosure:** J. Bowers. is a co-founder and shareholder of Nexus Photonics (I) and Quintessent (I), start-ups in silicon photonics.

**Data availability.** Data underlying the results presented in this paper are available.